\documentclass[times,twocolumn,final]{elsarticle}
\usepackage{amssymb}
\usepackage{latexsym}

\usepackage{url}
\usepackage{xcolor}
\usepackage{caption}
\usepackage{hyperref}

\definecolor{newcolor}{rgb}{.8,.349,.1}

\journal{Medical Image Analysis}

\begin{document}

%\verso{Given-name Surname \textit{et~al.}}

\begin{frontmatter}

%\title{3D Siamese Neural Networks for Re-Identification and Growth Detection of Pulmonary Nodules without Registration}

\title{Re-Identification and Growth Detection of Pulmonary Nodules without Image Registration Using 3D Siamese Neural Networks}

%\author[1]{Given-name1 \snm{Surname1}\corref{cor1}}
%\cortext[cor1]{Corresponding author: 
%  Tel.: +0-000-000-0000;  
%  fax: +0-000-000-0000;}
%\author[1]{Given-name2 \snm{Surname2}\fnref{fn1}}
%\fntext[fn1]{This is author footnote for second author.}
%\author[2]{Given-name3 \snm{Surname3}}
%% Third author's email
%\ead{author3@author.com}
%\author[2]{Given-name4 \snm{Surname4}}
%\address[1]{Affiliation 1, Address, City and Postal Code, Country}
%\address[2]{Affiliation 2, Address, City and Postal Code, Country}

\author[1,2]{Xavier Rafael-Palou\corref{cor1}}
\cortext[cor1]{Corresponding author}
\ead{xavier.rafael@eurecat.org}

\author[3]{Anton Aubanell}
%\ead{ton.aubanell@gmail.com}

\author[1]{Ilaria Bonavita}
%\ead{ilaria.bonavita@eurecat.org}

\author[2]{Mario Ceresa}
%\ead{mario.ceresa@upf.edu}

\author[2]{Gemma Piella}
%\ead{gemma.piella@upf.edu}

\author[1]{Vicent Ribas}
%\ead{vicent.ribas@eurecat.org}

\author[2,4]{Miguel A. González Ballester}
%\ead{ma.gonzalez@upf.edu}

\address[1]{Eurecat Centre Tecnològic de Catalunya, eHealth Unit, Barcelona, Spain}
\address[2]{BCN MedTech, Dept. of Information and Communication Technologies, Universitat Pompeu Fabra, Barcelona, Spain}
\address[3]{Vall d'Hebron University Hospital, Barcelona, Spain}
\address[4]{ICREA, Barcelona, Spain}

\begin{abstract}
%%%
Lung cancer follow-up is a complex, error prone, and time consuming task for clinical radiologists. Several lung CT scan images taken at different time points of a given patient need to be individually inspected, looking for possible cancerogenous nodules. Radiologists mainly focus their attention in nodule size, density, and growth to assess the existence of malignancy. In this study, we present a novel method based on a 3D siamese neural network, for the re-identification of nodules in a pair of CT scans of the same patient without the need for image registration. The network was integrated into a two-stage automatic pipeline to detect, match, and predict nodule growth given pairs of CT scans. Results on an independent test set reported a nodule detection sensitivity of 94.7\%, an accuracy for temporal nodule matching of 88.8\%, and a sensitivity of 92.0\% with a precision of 88.4\% for nodule growth detection.

%%%%
\end{abstract}

\begin{keyword}
\MSC 41A05\sep 41A10\sep 65D05\sep 65D17
%% Keywords

Lung cancer\sep Nodule detection\sep Nodule growth\sep Transfer learning\sep Deep Learning
\end{keyword}

\end{frontmatter}

%\linenumbers

\section{Introduction}

%Lung Cancer
Lung cancer is the leading cause of cancer death, regardless of gender or ethnicity. Only 19\% of all people diagnosed with lung cancer will survive after 5 years, but this percentage improves dramatically when the disease is diagnosed at early stages \citep{noone2018seer}.

%Context
Small lung nodules are the most common expression of early lung cancer. Their variability in size, texture, and morphology makes it difficult to detect them even for clinical specialists. The use of thin-slice helical chest computed tomography (CT) together with the recommendations established by clinical guidelines, such as those of the Fleischner Society \citep{macmahon2017guidelines}, has allowed improving nodule detection rates as well as better identifying the malignancy of incidental nodules. However, recommendations for borderline and complex cases are still vague and open to the judgment and experience of the clinicians.

Current clinical criteria for assessing pulmonary nodule changes rely on visual comparison and diameter measurements from the axial slices of the initial and follow-up CT images \citep{larici2017lung}. Three-dimensional assessment provides more accurate and precise nodule measurements, especially for small nodules \citep{ko2012pulmonary}. However, it requires the segmentation of the nodule, which is a time-consuming process and highly subjected to intra- and inter- observer variability. This is why it is rarely used in a typical clinical workflow.

%Add some paragraph/lines about follow up from the clinical perspective

%Current clinical problems 
Computer-assisted diagnosis (CAD) systems are expected to assist in clinical decision by providing relevant information such as accurate growth rates, increase in solid component, or change in density of the nodules. This information could help specialists to reduce the number of studies for a problematic nodule, decreasing the diagnostic time and, hopefully, reducing the classification of the neoplasm, which should lead to a reduction in morbid mortality \citep{american2014lung}.

%SOTA
Recent advances in deep neural networks \citep{goodfellow2016deep} have allowed increasing substantially the performances reported by conventional image processing methods in nodule detection \citep{setio2017validation}, segmentation \citep{messay2015segmentation}, and malignancy classification \citep{ciompi2017towards}. Some of the main advantages of using deep neural networks rely in their ability to learn and extract, in a very effective way, intricate patterns from the raw data without any previous feature engineering, reuse these patterns in different locations of the image, and even transfer them to different domains \citep{weiss2016survey}. Despite the recent explosion of methods based on deep neural networks in the lung cancer domain, most of them are focused on the analysis of a single CT scan. 

%Registration
Few CAD systems \citep{ardila2019end} have been proposed for the automatic support of lung cancer follow-up. Major developments in the field are mainly limited by the lack of open datasets with annotated series of CTs. To analyze series of CT scans, prior and follow-up lung exams have to be initially registered to facilitate, for instance, the correct re-identification of pulmonary nodules. Several factors compromise the effectiveness of the registration process, such as the variability in the image size and resolution originated by the use of different CT scans, and the variability in the position and breath cycle of the patients when performing the scanning. %Current sophisticated registration techniques rely on smooth deformations of the image using complex and hand-crafted engineered features extracted from the image  \citep{zikri2019toward}. 

Although current medical image registration methods \citep{song2017review}, especially non-linear \citep{ruhaak2017estimation}, report accurate CT alignments, they are still slow and introduce some distortions in the intrinsic structure of the lung, hindering their wide clinical acceptance \citep{VIERGEVER2016140}. In addition, other complexities must be addressed, regardless of the quality of the image registration, to enable a proper nodule re-identification, such as the existence of several nodules close to each other, and/or the alteration in texture, size, and even location of the nodules due to disease progression. Therefore, more research is still needed to reliably include the nodule re-identification in different CT scans, in automated tools to support physicians in the analysis of longitudinal studies of lung cancer.

%Novelty & main contributions
This work aims to take a step in this direction, and proposes a novel approach for the re-identification of pulmonary nodules. In particular, we propose a 3D Siamese neural network \cite{koch2015siamese} to predict the most likely matching nodules from a series of lung CT scans of the same patient. This approach does not require prior registration of the CT scans, avoiding some of the shortcomings that it entails. In addition, to demonstrate the value of this approach, we integrate it into an automated pipeline aimed to detect the growth of pulmonary nodules over time.

The contributions of this paper with respect to previous works is two-fold. First, we investigate and provide several models for re-identifying lung nodules in CT scans series, relying directly on 3D volumetric data, transfer learning, and siamese neural networks. In this sense, to the best of our knowledge, this would be the first time that the problem of pulmonary nodule re-identification is addressed through deep learning techniques. Secondly, we build and evaluate an automatic pipeline that integrates the proposed models to predict nodule growth from logitudinal CTs.

\section{Related work}

 %Current clinical criteria for assessing pulmonary nodule changes rely on visual comparison and diameter measurements from the axial slices of the initial and follow-up CT images \citep{larici2017lung}. Three-dimensional assessment provides more accurate and precise nodule measurements, especially for small nodules \citep{ko2012pulmonary}. However, it requires the segmentation of the nodule, which is a time consuming process and highly subjected to intra- and inter- observer variability. This is why it is rarely used in a typical clinical workflow.

%Despite these clear benefits, CNN require high amounts of data in order to provide accurate predictive models. In this regard, transfer learning [] has allowed building effective models without the need for large datasets by reusing previously trained models from other fields. Moreover, thanks to progress in more efficient hardware, research in deeper CNN architectures and new methods to train these models, 3D and even 4D raw data is being used directly to build these models rather than using multiple low dimensional views. This is specially convenient in the medical domain as most of the images are volumetric (such as CT, PET or MRI scans).%
\subsection{Automated nodule re-identification}

Lung nodule re-identification (i.e. matching) between current and former CT examinations is necessary for assessing nodule growth or shrinkage. While the majority of lung cancer CAD systems found in the literature focus on the nodule detection task \citep{loyman2019lung}, relatively few automated nodule matching systems have been proposed (partly because of the limited availability of follow up datasets). 

In \cite{lee2007performance} a matching rate of 67\% was obtained in a sample of 30 patients with metastatic pulmonary nodules. In screening datasets, higher matching rates are usually reported. In 54 pairs of low-dose multi detector CTs, a CAD system successfully matched 91.3\% of nodules $\geq$4mm \citep{beigelman2007computer}. In a low-dose multi detector CT screening study of 40 subjects with non-calcified nodules, a matching rate of 92.7\% across three time points was found \citep{tao2009automated}. Another CAD system \citep{koo2012improved} was evaluated for automated lung nodule matching using annotations from 4 experts in 57 patients. Performances obtained were between 79\% and 92\% of accuracy scores. These CAD systems relied on conventional computer vision techniques. Deep learning-based CAD systems for analysis of longitudinal lung cancer studies are practically nonexistent in literature. An exception is in \citep{ardila2019end}, where a CAD system for end-to-end lung cancer screening is proposed. However, nodule matching was not directly tackled in the study. 

All these CAD systems rely on registration of the lungs in the different examinations. Performing an accurate registration of lung images is particularly challenging due to the high deformability of the lung tissue and the volume changes during the breathing cycle \citep{murphy2010evaluation}. Several methods have been studied for the lung CT series analysis registration \citep{song2017review}. The choice of the right registration method and of the correct evaluation metric to assess its performance are of crucial importance as they can affect the results of the analysis. 

\subsection{Siamese Neural Networks}

The problem of nodule re-identification can be closely related to the one of recognizing the same object in different images. This type of problems has been successfully addressed by siamese neural networks \citep{bromley1994signature} (SNNs). They are designed as two sibling networks, connected by a distance layer at the top, trained to predict matching or mismatching between two input images. The original architecture, first introduced for the problem of signature verification, was later extended by \cite{koch2015siamese} using convolutional layers and adjusting the optimization metric with a weighted L1 distance between the twin feature vectors of both networks. 

SNNs have been extensively used in computer vision matching problems such as tracking objects in videos \citep{tao2016siamese}, matching pedestrians across multiple camera views \citep{varior2016gated}, and matching corresponding patches in satellite images \citep{hughes2018identifying}. 

In the medical image domain, SNNs have been used primarily to extract a latent representation for content-based image retrieval. For instance,  \cite{chung2017learning} proposed a SNN, pre-trained on the ImageNet dataset and using a contrastive loss function \citep{hadsell2006dimensionality} to retrieve similar images to the query, using a publicly available dataset of diabetic retinopathy fundus images. Another example is the work by \cite{cai2019medical}, which applied SNNs to retrieve similar images from several medical image databases of lung, pancreas, and brain. As far as we know, SNNs have not yet been applied to re-identify nodules in a series of lung CT scans.

%SNNs predict the similarity score between a couple of images based on pairwise difference computations. This function has been already reported \citep{staring2009image} to be a good indicator of the size and density change in lung nodules. Therefore, the use of SNNs could be advantageous to address at once the two problems of re-identification and quantification of pulmonary nodules.

\section{Method}

\subsection{Nodule re-identification}

To solve the problem of nodule re-identification in a pair of CTs of the same patient taken at different time points, we propose building a SNN \citep{koch2015siamese}. An appealing characteristic of SNNs is that they rely on a distance metric computed on features extracted automatically by a deep learning network. This should allow greatly accelerating and simplifying the nodule re-identification process avoiding to introduce a registration technique as source of variability and error in the analysis.

Siamese neural networks are composed of a feature extraction component in which two subnetworks (with shared architecture and weights) process a pair of images at a time to produce two embedding feature vectors directly from the images. A second component (i.e. the head of the network) aims to classify whether the two embedding feature arrays are similar or not. To assess this, the features are passed to a pairwise distance layer that computes a similarity score. 

In a previous study \citep{bonavita2019integration}, we trained a deep convolution neural network (CNN) for nodule classification able to effectively reduce the number of false positives in the nodule detection problem. In the present work, we have adjusted that network improving its final performance. In particular, we propose a 3D CNN based on a ResNet-34 architecture that expects nodule patches of 32x32x32. As described in the original paper, the patches are pre-processed crops done around the center of the annotated nodules of the lung CT. The nodule classification network was trained from scratch using a large amount of nodule candidates ($>$ 750K) from the LUNA-16 challenge dataset \citep{setio2017validation}. Further details on its architecture and performance are shown in the supplementary material (S1).

\begin{figure}[!ht]
\centering
\includegraphics[scale=.35]{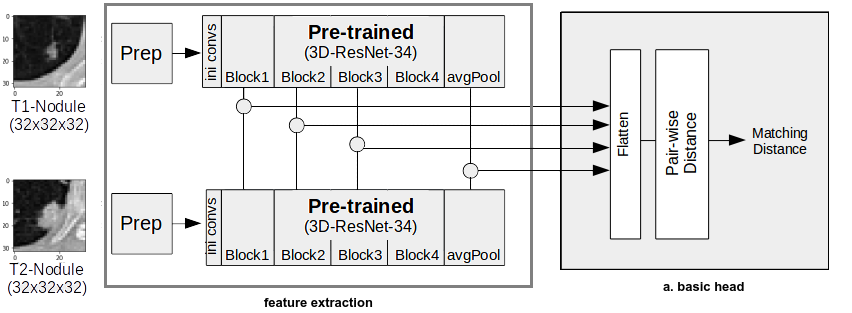}
\caption{Siamese network proposed for lung nodule re-identification. The network is composed of a feature extraction and a basic head network to perform the prediction.}
\label{fig:clasSiam}%
\end{figure}

In the current study, we removed the fully connected layers of the nodule classification network to use it as the backbone of the sibling networks of the feature extraction component of the SNNs. Figure-\ref{fig:clasSiam} shows the SNN architecture for the nodule re-identification problem. In this figure, we can observe the two components. First, the feature extraction component, which pre-processes the input nodule patches (i.e. taken at different time points, T1 and T2) and uses the sibling network to extract the corresponding feature maps. Second, the classification component composed of the head of the network that predicts if both feature maps are similar or not. These  feature maps (solid arrows in Figure-\ref{fig:clasSiam}) come from different levels of the pre-trained sibling networks. Further details about the feature maps and the network heads are described in Subsection 3.1.2 and 3.1.3, respectively.

\begin{figure}[!ht]
\centering
\includegraphics[scale=.35]{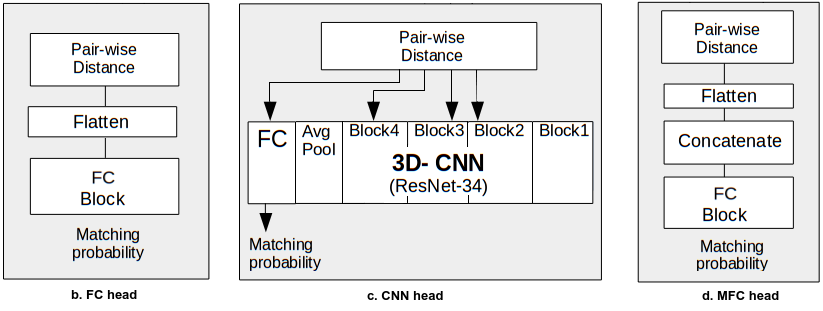}
\caption{Alternative head networks to configure different siamese networks.}
\label{fig:heads}%
\end{figure}

Different SNNs configurations were proposed (Table-\ref{tbl:siamesesConf}) to gain further insights into the best parameterizations. To allow a fair comparison of the configurations, we trained the SNNs with the same parameter values. Concisely, the number of epochs was set to 150, the learning rate to 0.0001, the batch size to 8, dropout to 0.3, the early stopping at 10 epochs without any significant improvement, and Adam \citep{kingma2014adam} was used for optimization. Finally, random rotation, flip, and zoom were applied for data augmentation.

\begin{table}[!ht]
\begin{tabular}{@{}lllll@{}}
\hline
  & Pre-trained  & Feature maps & Head & Loss  \\ 
\hline
FIBC  & Frozen & Individual & Basic & Contrastive \\ %FSDL
UIBC  & Unfrozen & Individual & Basic & Contrastive \\ %USDL
FIFB  & Frozen & Individual & FC & BCE\\ %FFCS
UIFB  & Unfrozen & Individual & FC & BCE\\ %UFCS
FICB & Frozen & Individual & CNN & BCE\\ %FCNNS
UICB & Unfrozen & Individual & CNN & BCE\\ %UCNNS
FCMB  & Frozen & Combined & MFC & BCE \\ %FMFS
UCMB  & Unfrozen & Combined & MFC & BCE \\ %UMFS
\hline
\end{tabular}
\caption{List of the different siamese network configurations. The column with acronyms is the result of joining the first letter of the options placed in the next 4 columns.}
\label{tbl:siamesesConf}
\end{table}

Below we describe in more detail the main configurations and parameters used in the experiments.

\subsubsection{Pre-trained network weights}

Two configuration values were proposed for this setting: frozen and unfrozen. Usually, the weights of the pre-trained networks in a SNN remain frozen. In this study the pre-trained network had a related but slightly different learning goal than the target (siamese) network. Thus, we allowed also the option of unfreezing the weights of the pre-trained network and updating them during the back-propagation steps of the siamese network training process. 

\subsubsection{Feature maps}

We proposed two options: using the feature maps individually and combining the feature maps together. Feature maps extracted from the first layers of a CNN refer to low-level and less domain-specific representations (e.g lines, circles, spikes), whereas features extracted from deeper layers are generally more high level and domain related representation (e.g. morphology, texture). To analyze the potential of both general and more specific nodule features, we propose to use features from different depths of the network (i.e., from the last layer of each of the 4 convolution blocks that holds the pre-trained Resnet-34 network). The resulting feature maps were obtained after a forward-passing through the network for each of the nodule images of the whole dataset. Table-\ref{tbl:features} shows the layer name, the number of filters per layer, the output dimension of each filter, and the total number of parameters for each of the selected feature maps.

\begin{table}[!ht]
\centering
\begin{tabular}{@{}llll@{}}
\hline
Layer  & Filters & Dimension & Total params  \\ 
\hline
layer1 & 64  & [16,8,8]   & 65536 \\
layer2 & 128 &  [8,4,4]   & 16384 \\
layer3 & 256 &  [4,2,2]   & 4096 \\
avgpool & 1  &  [1,1,512] & 512 \\
\hline
\end{tabular}
\caption{Layers selected from the pre-trained part of the SNNs.}
\label{tbl:features}
\end{table}

In total, we configured 4 experiments for the individual option (one for each of the 4 feature maps proposed), and 11 cases for the different possible combinations of the feature maps.

\subsubsection{Siamese heads}

We proposed four different head networks, one meant to follow a more conventional siamese architecture and the others with more exploratory purposes, more precisely:

\begin{enumerate}
    \item A basic head network (Figure-\ref{fig:clasSiam}) composed of a flatten (to homogenize all features to one dimension) and a pairwise distance (i.e. L1) layer, just after the feature extraction part of the network.
   \item A fully connected (FC) head network (Figure-\ref{fig:heads}b) composed of a pairwise distance, a flatten, and a FC block layer. The FC block comprises a FC layer (with 64 units), a batch norm, a ReLU, a dropout layer and a final FC layer (with one unit). This classifier head aims at finding non-linear patterns among the merged features (from both sibling networks).
    \item A CNN head network (Figure-\ref{fig:heads}c) composed of a pairwise distance layer and a clean (without pre-trained weights) ResNet-34 CNN. Several arrows connect the pairwise distance layer with this clean ResNet-34. There are as many arrows as pre-trained layers used to extract the features. The arrows redirect the features to a specific part of the clean ResNet-34. The redirection had to make compatible the dimensions of the output from the previous layer with the layers of the input. For instance, features extracted from last layer of block1 were linked to the initial layer of the block2, features from layer2 were linked to the initial layer of the block3 and so on. This head network aimed at exploring non-linear patterns between features but without loosing the space dimension (i.e. no flattening of the features was done between the pairwise layer and the clean ResNet-34).
    \item A multi-features combined (MFC) head network (Figure-\ref{fig:heads}d) composed of a pairwise distance layer, a flatten layer, a concatenation layer (to merge all features), and a FC (already described above). This head network aimed at exploring combination of features from different parts of the network.
\end{enumerate}

\subsubsection{Loss functions}

We explored two options: a contrastive loss and a binary cross entropy (BCE) loss function. Traditionally, SNNs are trained using a contrastive loss \citep{hadsell2006dimensionality} function. This function encodes both similarity and dissimilarity (between the feature maps) independently in a loss function. It ensures that semantically similar pairs are embedded close together while forcing the dissimilar pairs to be apart from each other. Another option to train these networks is through a prediction error-based approach. For our case we adopted the binary cross entropy loss. This implied to apply a sigmoid function on the outputs to transform them into probability values (between 0 and 1). 

\subsection{Nodule growth detection pipeline}

In order to evaluate our nodule re-identification approach in a more realistic and practical scenario, we integrated it into a pipeline that, given a pair of CTs of the same patient taken at two different time points (T1, T2), automatically assesses the nodule growth.

\begin{figure}[!ht]
\centering
\includegraphics[scale=.4]{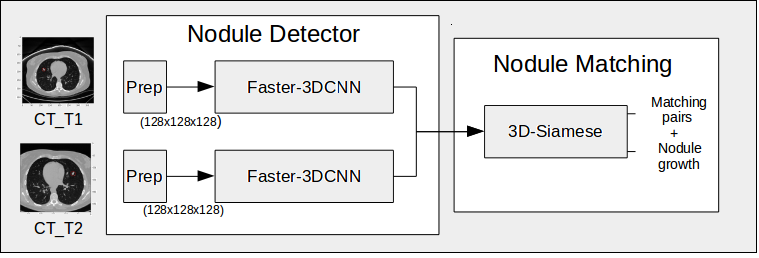}
\caption{Nodule growth detection pipeline.}
\label{fig:growthPipe}%
\end{figure}

The pipeline (Figure-\ref{fig:growthPipe}) comprises two components: 1) a nodule detector that, given a CT, generates a list of nodule candidates, and 2) a nodule matching component (embedding the siamese networks) that, given the list of nodule candidates of the CTs at the two time points, matches the nodules and computes the difference in diameter between them.

\subsubsection{Nodule detector}

To build the nodule detector, we followed the work of \cite{liao2019evaluate}, with which they won the Data Science Bowl lung cancer challenge\footnote{https://www.kaggle.com/c/data-science-bowl-2017}. The authors proposed a 3D Faster-RCNN \citep{ren2015faster} scheme for nodule detection. The backbone of the network was similar to the U-net \citep{ronneberger2015u} architecture, in which the information flows not only in a classical bottom-up way but also between the encoder and decoder parts of the network thanks to some symmetric links (or short-cuts) that bound both parts of the network. The output of this network were probability feature maps, useful for the lung cancer classification problem.

To the original network, we proposed attaching a double CNN head as in \citep{ren2015faster}. One head was used for regression and the other for classification. The regression branch infers the center (x,y,z locations) and the diameter of the nodule, while the classification branch predicts the probability of being a nodule.

The input lung CT was pre-processed before entering the nodule detection network. The image was resampled to an isotropic resolution (1x1x1mm), pixel intensities clipped between [-1000, 600] HU and normalized. The image was then split in overlapping patches (due to memory constraints) of 128x128x128 with an overlap of 32 pixels per dimension. Following \cite{liao2019evaluate}, each patch was fed to the network together with a second input of size  (32x32x32x3) which contains the relative locations of the patch image with respect to the whole scan. The final network architecture used for nodule detection as well as the performance obtained in LUNA-16 \citep{setio2017validation} dataset can be found in the supplementary material (S2).

\subsubsection{Nodule matching}

This component performs the re-identification of the nodules among all CT pairs. To do this, for each pair of CTs, we took each candidate found at T1 and we paired with each of the candidates found at T2. The pairs were pre-processed following the specifications described in Section 4.1, and then they were fed to the SNN. The network, trained off-line, provided a matching probability for each pair of candidates. The pairs with the highest probability were selected as the matching ones.

To assess the performance of this process, we computed for each pair of CTs, whether the candidate at T2 predicted with highest probability by the SNN, matched with the annotated nodule at T2. Additionally, we computed the time required for finding the matching nodules. We repeated this process for each of the SNN configurations.

Once having predicted all matching nodules for each pair of CTs, the pipeline returns the nodule growth along with the location and diameter of the matching nodules. The nodule growth is calculated directly by the difference between the predicted nodule diameters at T1 and T2 for each pair of lung CTs. 

To evaluate the nodule growth detection, we selected all the correctly matched CT pairs and compared whether the nodule growth difference was of the same sign in both ground truth and predicted. True positive (TP) and false negative (FN) cases were those that had (in both ground truth and predicted) positive and negative growth differences, respectively.  A false positive (FP) case was considered when the predicted growth difference was positive and the ground truth one was negative; and a false negative (FN) was considered in the opposite case. 

%\begin{figure}[!ht]
%\centering
%\includegraphics[scale=.5]{img/CT_analysis_new.png}
%\caption{Overall perspective of the proposed method.}
%\label{fig:method}%
%\end{figure}

\section{Experiments and results}

\subsection{Evaluation datasets}

\subsubsection{LUNA-16}

In this work we used an updated version of the LIDC dataset \citep{armato2015data} provided in the LUNA16 challenge \citep{setio2017validation}, which includes only scans with at least one lesion of size $\geq$ 3 mm marked as a nodule by at least three of the four radiologists. The LUNA16 dataset consists of 888 CT scans comprising a total of 1186 nodules. Annotations with coordinates of each nodule in the three spatial axes inferred from the original LIDC annotations are also provided. 

\subsubsection{VH-Lung}

This dataset was designed specifically to identify and follow up lung nodules in time. Ethics approval was obtained from the Medication Research Ethics Committee of
Vall d'Hebrón University Hospital (Barcelona) with reference number
PR(AG)111/2019 presented on 01/03/2019.

Inclusion criteria were patients without a previous neoplasia, with a confirmed diagnosis, and with visible nodules ($\geq$ 5 mm) in at least two consecutive CT scans separated in time by more than six months. These nodules were located in the three spatial axes by two different specialists at each time point and quantified by another experienced radiologist.

% Ethical committee approval
In total the dataset contains 151 cases with two thoracic CT scans. For each case, the clinicians annotated only one relevant nodule in both CT scans. The dataset was divided into two subsets, one for training (113 patients) with 70 cancer and 43 benign cases, and other for testing (38 patients) with 25 cancer and 13 benign cases. 

\subsection{Nodule re-identification} 

To train the different SNNs, we built a new balanced dataset from the  VH-Lung including 302 instances. As positive cases (N=151, label=1), we used the annotated locations of the matching nodules at T1 and T2. As negative cases (N=151, label=0), we used the nodule locations at T1 together with a random nodule location of the annotated nodule locations at T2 (avoiding to select correct nodule location). 
Random stratified sampling was used to partition the data into training (75\% of whole data) with 212 CT pairs (113 positive matching nodules) and testing sets, 90 CT pairs (with 38 positive cases). 

We optimized the different SNNs (Table-\ref{tbl:siamesesConf}) with the training data using a stratified 10-fold cross-validation, and we tested them with the testing set. Results of the best SNNs configurations are shown in Table-\ref{tbl:siameses}.

\begin{table*}[!ht]
\begin{tabular}{@{}lllllll@{}}
\hline
Configuration  & Layer & tr\_acc & val\_acc & test\_acc & test\_prec & test\_rec  \\ 
\hline
FIBC  &  layer2 & 0.790 $\pm$ 0.013 &  0.775 $\pm$ 0.051 & 0.709 $\pm$ 0.003 & 0.806 $\pm$ 0.002 & 0.550 $\pm$ 0.007 \\
UIBC  &  layer3 & 0.891 $\pm$ 0.009 &  0.864 $\pm$ 0.044 & 0.798 $\pm$ 0.018 & 0.765 $\pm$ 0.024 & 0.863 $\pm$ 0.036\\
FIFB  &  layer1 & 0.939 $\pm$ 0.025 &  0.899 $\pm$ 0.039 & 0.921 $\pm$ 0.036 & 0.905 $\pm$ 0.054 & 0.944 $\pm$ 0.038 \\
UIFB  &  layer2 & 0.918 $\pm$ 0.037 &  0.890 $\pm$ 0.039 & 0.896 $\pm$ 0.028 & 0.871 $\pm$ 0.050 & 0.934 $\pm$ 0.017\\
FICB &  layer1 & 0.867 $\pm$ 0.039 &  0.857 $\pm$ 0.060 & 0.831 $\pm$ 0.041 & 0.824 $\pm$ 0.075 & 0.860 $\pm$ 0.061\\
UICB &  layer1 & 0.868 $\pm$ 0.063 &  0.888 $\pm$ 0.049 & 0.859 $\pm$ 0.070 & 0.842 $\pm$ 0.093 & 0.900 $\pm$ 0.046 \\
FCMB  &  layer1, layer2 &  0.938 $\pm$ 0.034 & 0.882 $\pm$ 0.037 & 0.918 $\pm$ 0.017 & 0.907 $\pm$ 0.029 & 0.934 $\pm$ 0.035\\
UCMB &  layer1, layer2, avgpool &  0.954 $\pm$ 0.023 & 0.897 $\pm$ 0.045 & 0.925 $\pm$ 0.025 & 0.904 $\pm$ 0.040 & 0.952 $\pm$ 0.032\\
\hline
\end{tabular}
\caption{Performance results of the different SNN configurations}
\label{tbl:siameses}
\end{table*}

\subsection{Nodule growth detection pipeline}

\begin{figure}[!ht]
\centering
\includegraphics[scale=.35]{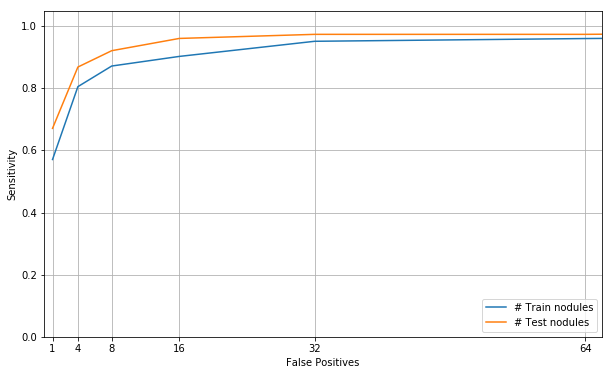}
\caption{FROC-curve of the malignant nodule detection algorithm for train and test partition.}
\label{fig:noddet}%
\end{figure}

To evaluate the performance of the nodule growth detection pipeline, we used the VH-Lung test set. First, we analyzed whether the relevant nodules (one per CT) annotated by the radiologists were correctly found. As the nodule detector outputs several candidates with an associated nodule probability, we explored the minimum number of candidates that allowed detecting the maximum number of annotated nodules. Thus, we computed a FROC-curve \citep{setio2017validation} for both train and test datasets to inspect the sensitivity of finding the (only) annotated nodule per scan at different FP rates. As we can observe in Figure-\ref{fig:noddet}, the model achieves high sensitivity scores with very few FP. The FROC-curve allowed us to set the threshold at 32 FPs, with only 11 annotated nodules out of 226 missed in the train set, which represented a sensitivity of 0.9513. In the test set (at 32 FP as threshold) only 2 out of 76 nodules were missed, resulting in a sensitivity of 0.9736.

\begin{figure}[!ht]
\centering
\includegraphics[scale=.5]{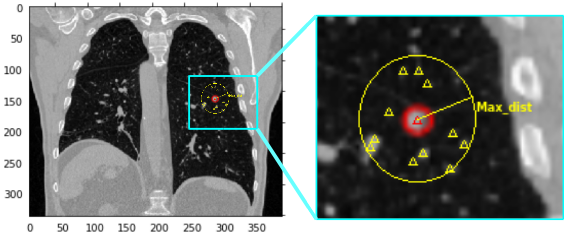}
\caption{Candidates predicted (i.e. yellow marks) at a maximum distance from the ground truth centroid (i.e. red circle).}
\label{fig:centroid}%
\end{figure}

To gain insight into the complexity of the re-identification problem, we computed how many candidates were located within a chosen Euclidean distance from the nodule ground truth position (Figure-\ref{fig:centroid}). We defined 5 different distance thresholds: radius squared Euclidean distance (as used in the LUNA-16 challenge to accept a nodule detection as correct) and 4 fixed Euclidean distances (30, 20, 15 and 10 mm). For every distance, we computed the number of CTs in which 0, 1, 2, 5 or more than 10 candidates fell within the distance. Moreover, we computed an accuracy of detection for every distance choice by dividing the number of CTs for which only one candidate is within the distance by the total number of CTs. Results are shown in Table-\ref{tbl:alignments}. 

\begin{table}[!ht]
\begin{tabular}{@{}lllllll@{}}
\hline
Distances  & N=0 & N=1 & N=2 & N=5 & N$\geq$10 & Ac \\ 
\hline
r**2  & 0 & 18 & 6 & 2 & 3 & 0.50\\
30 mm & 1 & 22 & 7 & 1 & 0 & 0.61\\
20 mm & 1 & 26 & 6 & 0 & 0 & 0.72\\
15 mm & 1 & 32 & 3 & 0 & 0 & 0.88\\
10 mm & 1 & 34 & 1 & 0 & 0 & 0.94\\
5 mm & 3 & 33 & 0 & 0 & 0 & 0.91\\
3 mm & 5 & 31 & 0 & 0 & 0 & 0.86\\
1.5 mm & 18 & 18 & 0 & 0 & 0 & 0.50\\
\hline
\end{tabular}
\caption{Detected candidates (N) per each CT at T2 at different Euclidean distances}
\label{tbl:alignments}%
\end{table}

Next, we evaluated the performance of the best SNN (Table-\ref{tbl:siameses}) for nodule re-identification using the location of the nodule candidates provided by the nodule detector. The best results were achieved by the FIFB network with only 4 CT-pairs incorrectly matched and an accuracy of 0.888. All results are presented in Table-\ref{tbl:matchingSNN}.

\begin{table}[!ht]
\begin{tabular}{@{}lllllll@{}}
\hline
Configuration  & Correct & Incorrect & Accuracy & Time(s) \\ 
\hline
FIBC  & 25 & 11 & 0.694  & 18.71\\
UIBC  & 27 &  9 & 0.750  & 36.018 \\
FIFB  & 32 &  4 & 0.888  & 9.36\\
UIFB  & 30 &  6 & 0.834  & 12.73\\
FICB & 30 &  6 & 0.834  & 20.12\\
UICB & 28 &  8 & 0.777  & 20.16\\
FCMB  & 31 &  5 & 0.861  & 12.41\\
UCMB  & 31 &  5 & 0.861  & 19.10\\
\hline
\end{tabular}
\caption{Results of the different nodule re-identification pipelines}
\label{tbl:matchingSNN}%
\end{table}

\begin{figure}[!ht]
\centering
\includegraphics[scale=.6]{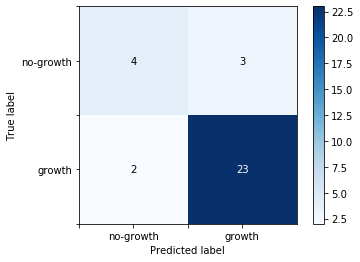}
\caption{Confusion matrix for nodule growth prediction}
\label{fig:cm_growth}%
\end{figure}

Then, we evaluated the performance of the best pipeline (i.e. the pipeline configured with the FIFB network) for the nodule growth detection task. As explained in Section 4.2.2, a correct prediction was achieved when the difference on diameters between predicted and ground truth nodules had both the same sign. In this way, having 32 correctly identified cases (out of 36), we obtained a 0.92 of recall, a 0.88 of precision and a 0.90 of F1-score. The confusion matrix is shown in Figure-\ref{fig:cm_growth}.

\begin{figure}[!ht]
\centering
\includegraphics[scale=.35]{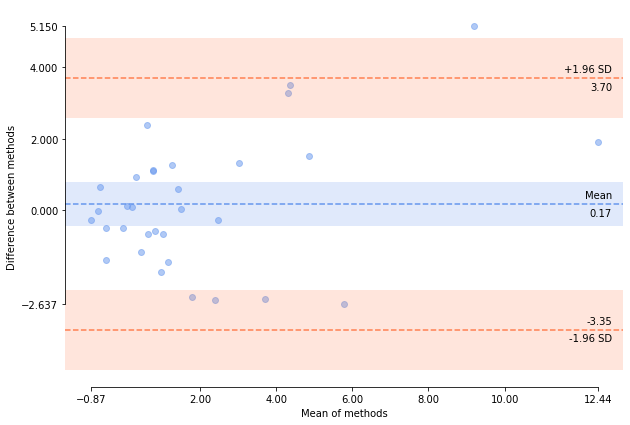}
\caption{Bland–Altman plot for agreement between predicted and ground truth nodule growth}
\label{fig:blandAlt}%
\end{figure}

\begin{figure*}[!ht]
\centering
\captionsetup{justification=centering,margin=2cm}
\includegraphics[width=\textwidth]{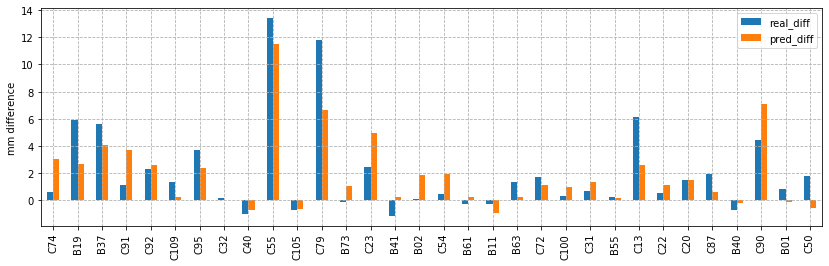}
\includegraphics[scale=.6]{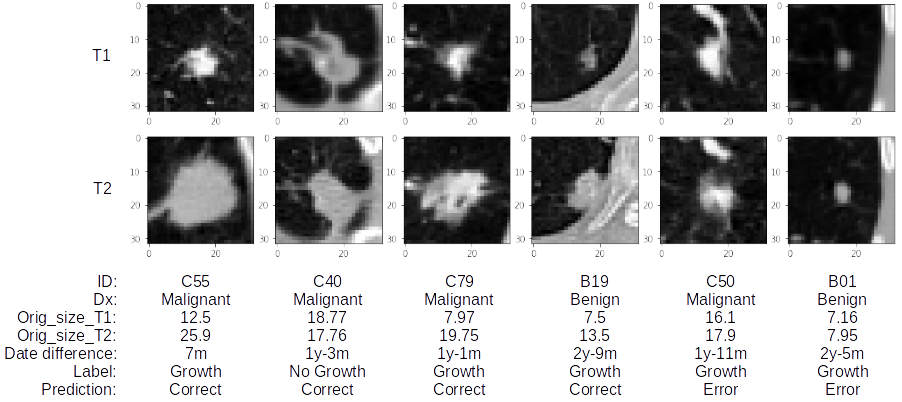}
\caption{Comparison between real and predicted cases. Upper panel: diameter differences for all test set. Lower panel: axial slices at two time points of different nodules.}
\label{fig:diffdiam}%
\end{figure*}

Additionally, we assessed the quality of the diameter measure prediction. Agreement between the predicted and ground-truth nodule growth vectors was assessed with a Bland-Altman \citep{altman1983measurement,jaketmp_2018_1256204} plot (Figure-\ref{fig:blandAlt}). The mean difference between the two measurements was 0.17 mm with a 95\% confidence interval (from -3.35 to 3.70mm). The mean value of the difference was not significantly different from 0 on the basis of a 1-sample t-test (p-value = 0.99), for which a previous logarithm transformation and a data inspection was carried out to ensure the prerequisite assumptions of normality. Also, we computed the mean absolute error of the predicted nodule growths (1.38 $\pm$ 1.17 mm), their mean squared error (3.26 $\pm$ 5.30 mm) and its coefficient of determination (r$^2$=0.71). Finally, Figure-\ref{fig:diffdiam} shows the predicted and real difference of diameters for all CT pairs of the test dataset. To support the interpretation of this figure, we have included the axial slice with major diameter taken at time points T1 and T2 of an illustrative subset of nodules.

\section{Discussion}

%Summary of the method
In this article, we provide a novel way to address the nodule re-identification problem. In particular, we propose a deep SNN that can directly re-identify nodules located in a series of pairs of CT scans without the need for any image registration. %Some of the reasons behind this choice are that deep neural networks are especially good for finding good descriptors of an image, and SNNs have been proven to be successful architectures for solving the problem of finding similar objects in different images \citep{taigman2014deepface}.

% Main approach and how it can overcome previous limitations and its benefits
The SNN allows matching pulmonary nodules in different CTs in a single stage by outputting a similarity score (i.e. the probability of being the same nodule). In contrast, standard techniques require at least two stages: first registering the image and then identifying matching nodules with some distance function. Moreover, with the proposed solution, no additional deformations/perturbations of the lung scan are performed, so that nodule measurements can be done directly from the image itself. Another advantage is that the re-identification process is fast since all weights of the network have already been calculated during the training phase.

%Matching results: training
We designed and tested several SNN architectures in order to fully understand the complexities of the problem and find the best network configuration. Results (Table-\ref{tbl:siameses}) show that, in general (7 out of 8 experiments), the  networks obtained high accuracy scores, above 85\% in validation and 80\% in test. Indeed, several of the SNN configurations (e.g. FIFB, UCMB) achieved accuracy scores in test above $92\%$, in agreement with the state-of-the-art performances reported for automated matching of pulmonary nodules \citep{tao2009automated, koo2012improved}. One of the main factors contributing to the good performance is the use of transfer learning, namely initializing the backbone of the different SNNs with the weights of a previously trained 3D network. This can explain why, even the simpler network (FIBC), without the addition of extra layers and without any fine adjustments, achieved nevertheless good performances (77.5\% in validation and 71\% in test).

Regarding the loss functions configured in the different experiments, the methods using the BCE loss (which are based on probabilities) slightly outperformed the ones using the contrastive loss (which is based on distances). This can be seen in the difference in accuracy (3.5\% in validation and 12\% in test) obtained by the best network configured with probability-based loss function (FIFB) compared to the best network configured with loss function based on distance (UIBC).

Another finding was that unfreezing the weights of the pre-trained networks usually allowed for better performances. This is particularly evident in the UIBC case, which exceeded of almost 10\% in validation and testing the corresponding frozen configuration (FIBC). Somehow, this finding was expected as weights were transferred from networks trained in a different, although closely related, domain.

With respect to the features used by the networks, we can observe (Table-\ref{tbl:siameses}) that, in almost all the methods, the best performance was achieved by using features extracted by layer1 and/or layer2, while only for two methods it was achieved using features from layer3 and avgpool (i.e. the global average pooling). This may suggest that features encoding simple patterns (from earlier layers) are preferred for this problem, whereas layers that contains more specific features (from the last layers) are less useful. It is also worth noticing that networks combining features from different layers did not clearly outperform networks using features from a single layer. This is  the case of UCMB in which the reported validation performances are just a bit lower (0.2\%) than the performances reported by the FIFB configuration, although in test, UCMB outperformed by 0.4\% the performance of FIFB.

Concerning the type of heads with which the networks were configured, the best option was using fully connected layers (FC head). Surprisingly, networks with extra convolution layers before the fully connected layers (CNN head) achieved worse performances (1\% and 6\% less in validation and test, respectively) than networks with FC heads. This might suggest that adding extra convolution layers to find patterns between locally connected features increases the complexity of the model, leading to more weights to adjust but with the same amount of training data.

%Matching results: pipeline 
To test these networks in a more realistic scenario, we integrated them into automatic two-stage pipelines aimed at first re-identifying pulmonary nodules and then predicting nodule growth given series of CTs of the same patient. In this setting, the performances achieved by the different nodule re-identification networks (Table-\ref{tbl:matchingSNN}) were a bit lower than in training phase. This may be due to the fact that the patched images were cropped around the position predicted by the nodule detector, as opposed to in training which they were cropped around the centroid of the nodules. This may have occluded some parts of the nodules, making its correct matching more difficult. However, 5 out of 8 networks reached a nodule matching accuracy score above 80\%. As in training, the network with the best performance was the FIFB, with a 88.8\% of accuracy. 
This performance score is slightly worse than current state-of-the art ($>92\%$). However, we must highlight that, in our approach, the position of the nodules were automatically provided by the nodule detector (without any prior human intervention), whereas in \citep{tao2009automated, koo2012improved} the position of the reference nodule to match is given by the radiologist.

%We highlight this result taking into account that the nodules were automatically re-identified (i.e. without any specialist intervention), unlike previous reported studies \citep{tao2009automated, koo2012improved} where the reference nodules were marked manually by the radiologists. 

%Time matching
In terms of computational time, our approach achieved satisfactory performances being able to re-identify the nodules of the complete test set, in times ranging from half a minute (in the worst case, UIBC) to less than 10 seconds (for the best configuration, FIFB), as can be seen in Table-\ref{tbl:matchingSNN}.  This is a particularly appealing feature of our method, since even the most recent techniques for registration of lung CT images, necessary by any standard pipeline for nodule re-identification, require significantly more time, for instance 5 minutes according to \cite{ruhaak2017estimation} or approximately 1 minute by \cite{zikri2019toward} per case. These processing times fluctuate substantially depending on the technique and the quality of the image registration.

%recent performance of rigid, affine, and deformable intensity- and feature-based registration on lesion-centered ROIs \citep{zikri2019toward} requires from 0.7 to 17.2 minutes per CT.

%Nodule growth
Although the focus of the paper is the nodule re-identification, to explore the potential applications of the method, we also quantified and assessed nodule growth. To do this, we used the best network for nodule re-identification (FIFB) and integrated it in the nodule-growth pipeline. In total, nodule growth was correctly detected in 27 cases and erroneously in 5 cases. However, only 2 of these errors were false negatives (that is, the pipeline failed to predict growth); one of them was on a benign nodule (B01) with growth difference of less than 1 mm, whereas the other was on a malignant nodule (C50) with growth difference of 1.8 mm. As shown in Figure-\ref{fig:blandAlt}, there is an agreement when comparing predicted and real nodule growths as most of the measures fall between the two standard deviations of the mean, there is a non-significant difference between them (p=0.99), and they show a good correlation score (r$^2$=0.71). Despite this positive results, the values obtained for the 95\% limits of agreement ($>$ 3mm) are still high. Thus, emphasis should be done in improving the current nodule detector to provide more accurate diameters. Nevertheless, from a clinical point of view, the majority of the nodule differences were correctly classified (growth, no-growth) as shown in Figure-\ref{fig:cm_growth}, and we reported a mean absolute error of 1.38 $\pm$ 1.17mm in diameter with respect to the ground truth, which indeed is slightly less than the 1.73 and 2.2mm of the variability error reported in different retrospective analysis \citep{revel2004two, kim2016measurement} measuring changes in solid and subsolid nodules ($<$2cm) using only their diameter.

%In order to obtain a better assessment of the nodule-growth on the VH-Lung dataset through the pipeline, a straightforward improvement could be retraining the nodule detector on the VH-Lung dataset -remember it was originally trained on a different dataset (LUNA-16). Improving the performance of the nodule detector, would allow a more precise location and diameter quantification of the nodules, directly impacting the performance of nodule re-identification and growth detection.

%Limitations
This study, however, is subject to several limitations. In the medical domain, data is a scarce and difficult resource to obtain. Having an insufficiently large dataset can negatively impact the performance of deep learning-based models. This is even more concerning for re-identification of lung nodules, since for each patient, twice as many images and annotations are needed. Another main limitation of the study is that the only expert annotation provided for nodule quantification was the major axial diameter. Although the diameter is the most common radiological measure used in practice for nodule growth assessment, using 3D measurements could lead to a more accurate quantification. In addition, if we would have had nodule measurements from more experts, we could have better explained the clinical variability, reporting more accurately the performance of our pipeline with respect to nodule growth prediction. Finally, in this work, we focus on training and evaluating several SNNs to explore different configurations. Finer tuning of hyperparameters (e.g. the learning rates, batch sizes or dropout values) may lead to improved results.

%Future work
Several future works are envisaged to extend the research presented in this paper. Applying different feature fusion techniques, introducing different manners to weigh the feature maps, applying new techniques to reduce the dimensionality of the problem, as well as the use of segmentation are just some of the research lines that can be explored beyond the presented work.

\section{Conclusions}
In this paper, we address the problem of automatic re-identification of pulmonary nodules in lung cancer follow-up studies, using siamese neural networks (SNNs) to rank similarity between nodules, which overpasses the need of image registration. This change of paradigm avoids possible image disturbances and provides computationally faster results. Different configurations of the conventional SNN were examined, ranging from the application of transfer learning, using different loss functions, to the combination of several feature maps of different network levels. The best results during the off-line training of the SNNs reached accuracies (0.89 in cross-validation and 0.92 in test) similar to those reported by state of the art registration mechanisms. Finally, we embedded the best SNN into a two-stage nodule growth detection pipeline. Nodule re-identification results reported by the pipeline in an independent test set were fast ($<$10 seconds, matching 38 pairs of CTs) and precise (0.88 accuracy score). Nodule growth predictions were also accurate (0.92 sensitivity score), and both the predicted and the ground truth measurements were not significantly different (p=0.99).

\section*{Acknowledgments}
This work was partially funded by the Industrial Doctorates Program
(AGAUR) grant number DI079, and the Spanish Ministry of Economy and Competitiveness (Project INSPIRE FIS2017-89535-C2-2-R, Maria de Maeztu Units of Excellence Program MDM-2015-0502).

%%Harvard
\bibliographystyle{model2-names.bst}\biboptions{authoryear}
\bibliography{refs}

\end{document}

%To number supplemental material with 'S':
\renewcommand{\thepage}{S\arabic{page}} 
\renewcommand{\thesection}{S\arabic{section}}  
\renewcommand{\thetable}{S\arabic{table}}  
\renewcommand{\thefigure}{S\arabic{figure}} 

\title{Supplementary material}
\author{}
\date{}

\maketitle

%\section{The malignancy classifiers}

%\begin{figure}[htb!]
%\centering
%  \includegraphics[width=0.95\textwidth]{img/malignancy_classifiers_arch.png}
%  \caption{Architecture of the two CNN networks used.}
%  \label{fig:cnn_blocks}
%\end{figure} 

\section{Nodule classification}

The model implemented for nodule classification is a 3D ResNet-34, borrowed from \citep{hara3dcnns}. We used this architecture rather than the ResNet-50 described in our previous work \citep{bonavita2019integration} because we achieved better sensitivity and FROC scores. To train the network, we used the Adam optimization algorithm, a batch size of 128, a weighted binary cross entropy loss function (to attenuate the heavy data imbalance) and 3D data augmentation (flip, rotation, lighting and zoom transforms).  

\subsection{Results}

Tables (\ref{tbl:res_cm}, \ref{tbl:res_fp}) describe the evaluation results of the network in the test set of the LUNA-16. This partition represented 10\% of the total amount of data, and It was composed of 75780 candidates (labeled as 0) and 144 nodules (labeled as 1).

\begin{table}[ht!]
\centering
\begin{tabular}{llll}
 & & \multicolumn{2}{l}{Predicted} \\ \cline{3-4} 
 & \multicolumn{1}{l|}{Real} & \multicolumn{1}{c|}{\begin{tabular}[c]{@{}c@{}}False\\ (0)\end{tabular}} & \multicolumn{1}{c|}{\begin{tabular}[c]{@{}c@{}}True\\ (1)\end{tabular}} \\ \cline{2-4} 
\multicolumn{1}{l|}{} & \multicolumn{1}{l|}{Candidate (0)} & \multicolumn{1}{l|}{75677} & \multicolumn{1}{l|}{103} \\ \cline{2-4} 
\multicolumn{1}{l|}{} & \multicolumn{1}{l|}{Nodule (1)}    & \multicolumn{1}{l|}{15} & \multicolumn{1}{l|}{129} \\ \cline{2-4} 
\end{tabular}
\caption{Confusion matrix results for the 3D ResNet network.}
\label{tbl:res_cm}
\end{table}

\begin{table}[ht!]
\centering
\begin{tabular}{l|l|l|l|l|}
\cline{2-5}
                                    & Precision & Recall & F1-score   & Support \\ \hline
\multicolumn{1}{|l|}{Candidate (0)} & 1.00      & 1.00   & 1.00 & 75780   \\ \hline
\multicolumn{1}{|l|}{Nodule (1)}    & 0.56      & 0.90   & 0.69 & 144     \\ \hline
\end{tabular}
\caption{Classification results for the 3D ResNet network.}
\label{tbl:res_fp}
\end{table}

\section{Nodule detector architecture}

\begin{figure}[htb!]
\centering
  \includegraphics[width=1.0\textwidth]{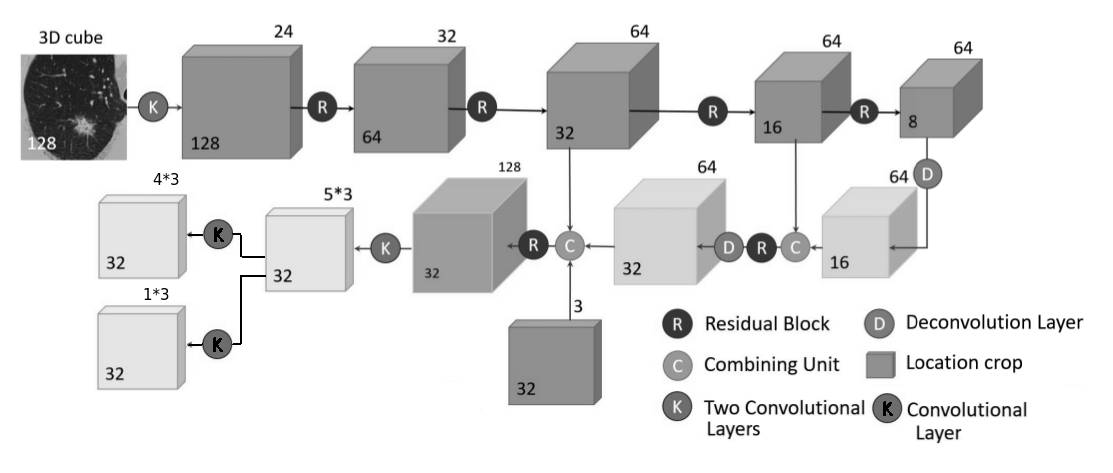}
  \caption{Architecture for the nodule detector.}
  \label{fig:nodDect1}
\end{figure}

The nodule detection network was implemented using the available code of two recent and successful works \citep{liao2019evaluate} and \citep{zhu2018deeplung}. The network was trained using a batch size of 8, Adam as optimization algorithm, and a learning rate of 0.1 with a decay of 0.001 every 100 epochs, with a total of 450 epochs. Moreover, we used hard negative mining \citep{shrivastava2016training} with a factor of 20 times the batch size, as well as random rotation, flip, and zoom for 3D data augmentation. The final network architecture is shown in figure-\ref{fig:nodDect1}. 

\subsection{Results}

\begin{figure}[htb!]
\centering
  \includegraphics[width=0.95\textwidth]{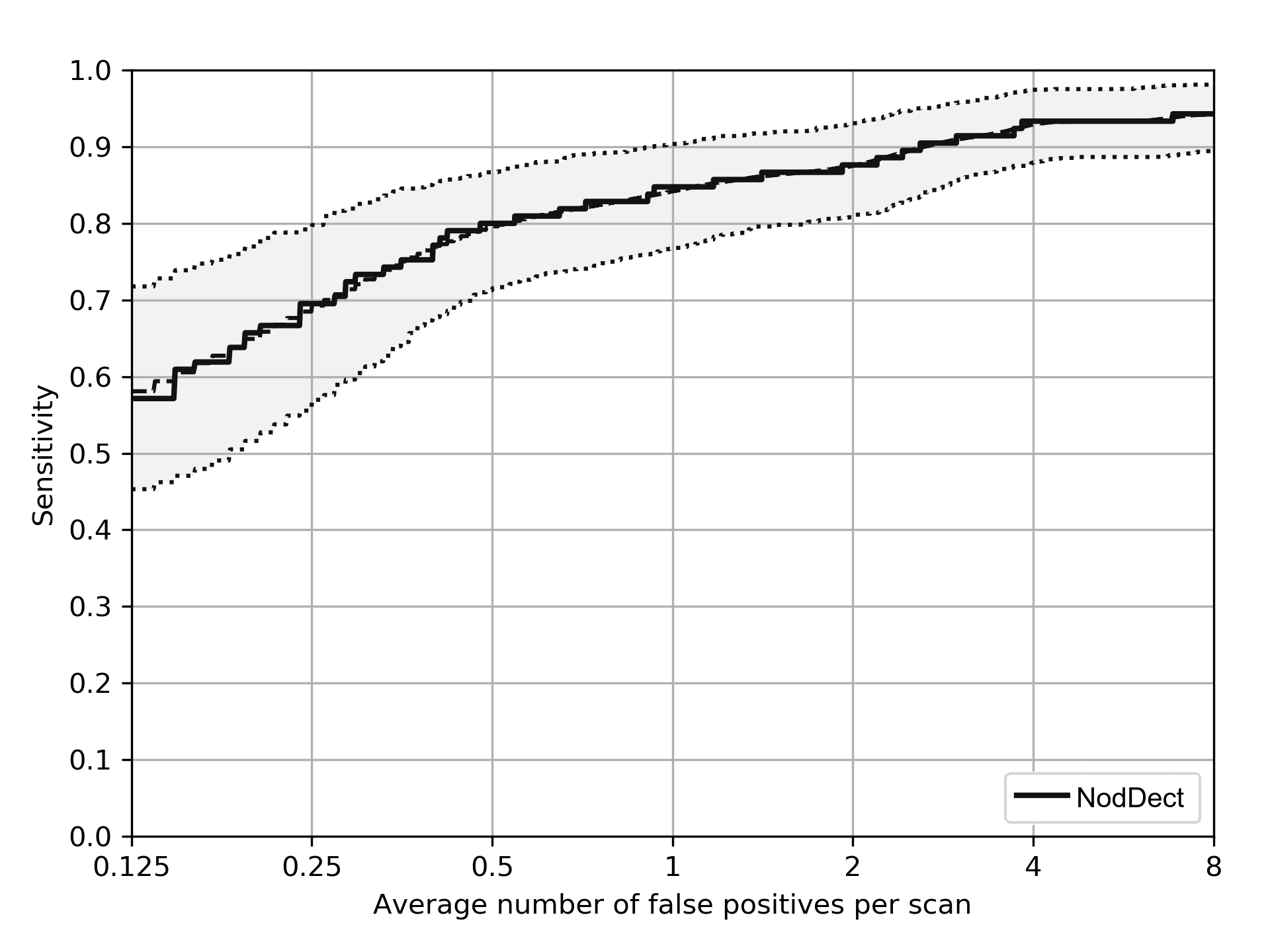}
  \caption{FROC curve of the lung nodule detector computed for the test set.}
  \label{fig:nodDectFROC}
\end{figure}

We evaluated the performance of this network with the test set partition of the LUNA-16 dataset. This partition was composed of 88 CT scans (out of 888 in total) with 105 annotated nodules. For each nodule annotation, we had its location (x,y,z) and its diameter. The resulting nodule detection performances are shown in Table-\ref{tbl:configNd}. The first column of the table has the different false positive (FP) ratios (averaged per scan), and the rest of columns show the sensitivity obtained (mean, upper, and lower bounds). Upper and lower bounds were obtained after a 1000 bootstrapping. Figure-\ref{fig:nodDectFROC} shows the FROC curve reported by this method.

\begin{table}[ht!]
\centering
\begin{tabular}{l|l|l|l|}
\cline{2-4}
FP\-Rate         & Mean & Lower & Upper \\ \hline
\multicolumn{1}{|l|}{0.125} & 0.5799 & 0.45217 & 0.7176       \\ \hline
\multicolumn{1}{|l|}{0.25}  & 0.6926 & 0.56976 & 0.7978       \\ \hline
\multicolumn{1}{|l|}{0.50}  & 0.7961 & 0.71544 & 0.8666       \\ \hline
\multicolumn{1}{|l|}{1.0}   & 0.8421 & 0.76800 & 0.9036       \\ \hline
\multicolumn{1}{|l|}{2.0}   & 0.8755 & 0.81132 & 0.9306       \\ \hline
\multicolumn{1}{|l|}{4.0}   & 0.9290 & 0.87962 & 0.9743       \\ \hline
\multicolumn{1}{|l|}{8.0}   & 0.9419 & 0.89423 & 0.9809      \\ \hline
\end{tabular}
\caption{Performances of the lung nodules detector at different FP in average per scan.}
\label{tbl:configNd}%
\end{table}

\bibliographystyle{model2-names.bst}\biboptions{authoryear}
\bibliography{refs}